\documentclass[12pt]{article}
\usepackage{graphicx}

\newcommand\pubdate{September 30, 2015}

\def\napoli{Institute for Nuclear Physics Mainz\\
Johannes-Gutenberg University, D-55128 Mainz, GERMANY}

\def\Title#1{\begin{center} {\Large #1 } \end{center}}
\def\Author#1{\begin{center}{ \sc #1} \end{center}}
\def\Address#1{\begin{center}{ \it #1} \end{center}}

\newcommand\pubblock{\rightline{\begin{tabular}{l} \\
         \pubdate  \end{tabular}}}
\newenvironment{Abstract}{\begin{quotation}  }{\end{quotation}}
\newenvironment{Presented}{\begin{quotation} \begin{center} 
             PRESENTED AT\end{center}\bigskip 
      \begin{center}\begin{large}}{\end{large}\end{center} \end{quotation}}

\def\beq{\begin{equation}}
\def\eeq#1{\label{#1}\end{equation}}
\def\eeqn{\end{equation}}

\def\beqa{\begin{eqnarray}}
\def\eeqa#1{\label{#1}\end{eqnarray}}
\def\eeqan{\end{eqnarray}}

\let\bar=\overbar

\def\Dslash{\not{\hbox{\kern-4pt $D$}}}
\def\dslash{\not{\hbox{\kern-2pt $\del$}}}

\def\msb{{\bar{\ssstyle M \kern -1pt S}}}

\begin{document}
\begin{titlepage}
\pubblock

\vfill
\Title{Measurement of the $e^+e^-\rightarrow\pi^+\pi^-$ Cross Section Using Initial State Radiation at BESIII}
\vfill
\Author{ Benedikt Kloss on behalf of the besiii collaboration}
\Address{\napoli}
\vfill
\begin{Abstract}
Using a data set with an integrated luminosity of 2.93 fb$^{-1}$ taken at a center-of-mass energy of 3.773 GeV with the BESIII detector at the BEPCII collider, we extract the $e^+e^-\rightarrow \pi^+\pi^-$ cross section and the pion form factor $|F_\pi|^2$ in the energy range between 600 and 900 MeV. We exploit the method of initial state radiation for this measurement, yielding a systematic uncertainty of 0.9\%. We calculate the contribution of the measured cross section to the leading-order hadronic vacuum polarization contribution to $(g-2)_\mu$. 
\end{Abstract}
\vfill
\begin{Presented}
CIPANP 2015\\
Vail, USA,  May 19 - 24, 2015
\end{Presented}
\vfill
\end{titlepage}
\def\thefootnote{\fnsymbol{footnote}}
\setcounter{footnote}{0}

\section{Introduction}

Precise measurements of hadronic cross sections  are an important input for the hadronic vacuum polarization contribution of the standard model (SM) prediction of the anomalous magnetic moment $(g-2)_\mu$~\cite{Jegerlehner}. Currently, a discrepancy of 3.6 standard deviations is found between the direct measurement of $a_\mu \equiv (g-2)_\mu/2$ and its SM prediction~\cite{g-2_strong_4}. The accuracy of the SM prediction of $(g-2)_\mu$, is entirely limited by the knowledge of the hadronic vacuum polarization contribution, which is obtained in a dispersive framework by using experimental cross section data $\sigma(e^+e^-\rightarrow \rm hadrons)$. The cross section $\sigma_{\pi\pi} = \sigma(e^+e^-\rightarrow \rm \pi^+\pi^-)$ contributes more than 70\% to this dispersion relation and, hence, is the by far most important exclusive hadronic channel of the total hadronic cross section. \\

\noindent The two most accurate measurements of $\sigma_{\pi\pi}$ have been obtained by the KLOE collaboration in Frascati \cite{2pi_KLOE05,2pi_KLOE08,2pi_KLOE10,2pi_KLOE12}, and the BABAR collaboration at SLAC \cite{2pi_BaBar_PRL,2pi_BaBar}. Both experiments claim an accuracy of better than 1\% in the energy range below 1 GeV, in which the $\rho(770)$ resonance is dominating the hadronic cross section. However, a discrepancy of approximately 3\% on the peak of the $\rho(770)$ resonance is observed. The discrepancy is even increasing towards higher energies and has a large impact on the SM prediction of  $a_\mu$. This shows the necessity of a reference experiment with a precision also in the order of 1\%.\\

\noindent This measurement can be done at the BESIII experiment, located at the symmetric $e^+e^-$ collider BEPCII in Beijing, China. Therefore, we analyze a data set of \mbox{2.93 fb$^{-1}$~\cite{2pi_BES}} taken at a center-of-mass (cms) energy $\sqrt{s}$ = 3.773 GeV. In this analysis we study the two-pion invariant mass range between 600 and 900 MeV/c$^2$. This range corresponds to the important $\rho$ peak region, which contributes more than 70\% to the two-pion contribution $a_\mu^{\pi\pi}$ and to about 50\% of the total hadronic vacuum polarization correction of $a_\mu$.

\section{Experiment}
The BESIII detector is located at the double-ring Beijing electron-positron collider (BEPCII)~\cite{BESIII}.
The cylindrical BESIII detector covers 93\% of the full solid angle. It consists of the following detector systems.
(1) A Multilayer Drift Chamber (MDC), filled with helium gas, composed of 43 layers, which provides a spatial resolution of 135 $\mu$m, an ionization energy loss $dE/dx$ resolution better than 6\%, and a momentum resolution of 0.5\% for charged tracks at 1~GeV/$c$.
(2) A Time-of-Flight system (TOF), built with 176 plastic scintillator counters in the barrel part, and 96 counters in the endcaps. The time resolution is 80 ps in the barrel and 110 ps in the endcaps. For momenta up to 1 GeV/$c$, this provides a 2$\sigma$ K/$\pi$ separation.
(3) A CsI(Tl) Electro-Magnetic Calorimeter (EMC), with an energy resolution of 2.5\% in the barrel and 5\% in the endcaps at an energy of 1 GeV.
(4) A superconducting magnet producing a magnetic field of 1T.
(5) A Muon Chamber (MUC) consisting of nine barrel and eight endcap resistive plate chamber layers with a 2 cm position resolution.


\section{Measurement}
We exploit the method of initial state radiation (ISR) for this measurement. One of the incoming beam particles radiates a high energetic photon and, thus, the available energy to produce the hadronic $\pi^+\pi^-$ final state is reduced and the two-pion mass range $m_{\pi\pi}$ below the cms energy becomes available. Thus, we select events of the type $e^+e^-\rightarrow\pi^+\pi^-\gamma$. Due to the geometrical acceptance of the BESIII detector only tagged events can be studied in the mass range between 600 and 900 MeV/c$^2$, where the ISR photon is explicitly measured in the EMC.\\

\noindent To suppress $e^+e^-\rightarrow \mu^+\mu^-\gamma$ events, the main background in this analysis, an artificial neural network (ANN) \cite{TMVA} was trained and tested using $\pi^+\pi^-\gamma$ and $\mu^+\mu^-\gamma$ Monte-Carlo (MC) samples as input. The {\sc phokhara} event generator \cite{Phokhara,Phokhara7}, interfaced with the \textsc{geant4}-based detector simulation~\cite{GEANT1,GEANT2}, was used to generate these kind of events. For a high precision analysis, possible discrepancies between data and MC simulation due to imperfections of the detector simulation, have to be considered. Track-based data-MC correction factors are obtained by comparing nearly background-free pion and muon samples in data and MC.\\

\begin{figure}[htb]
\centering
\includegraphics[width = 12cm]{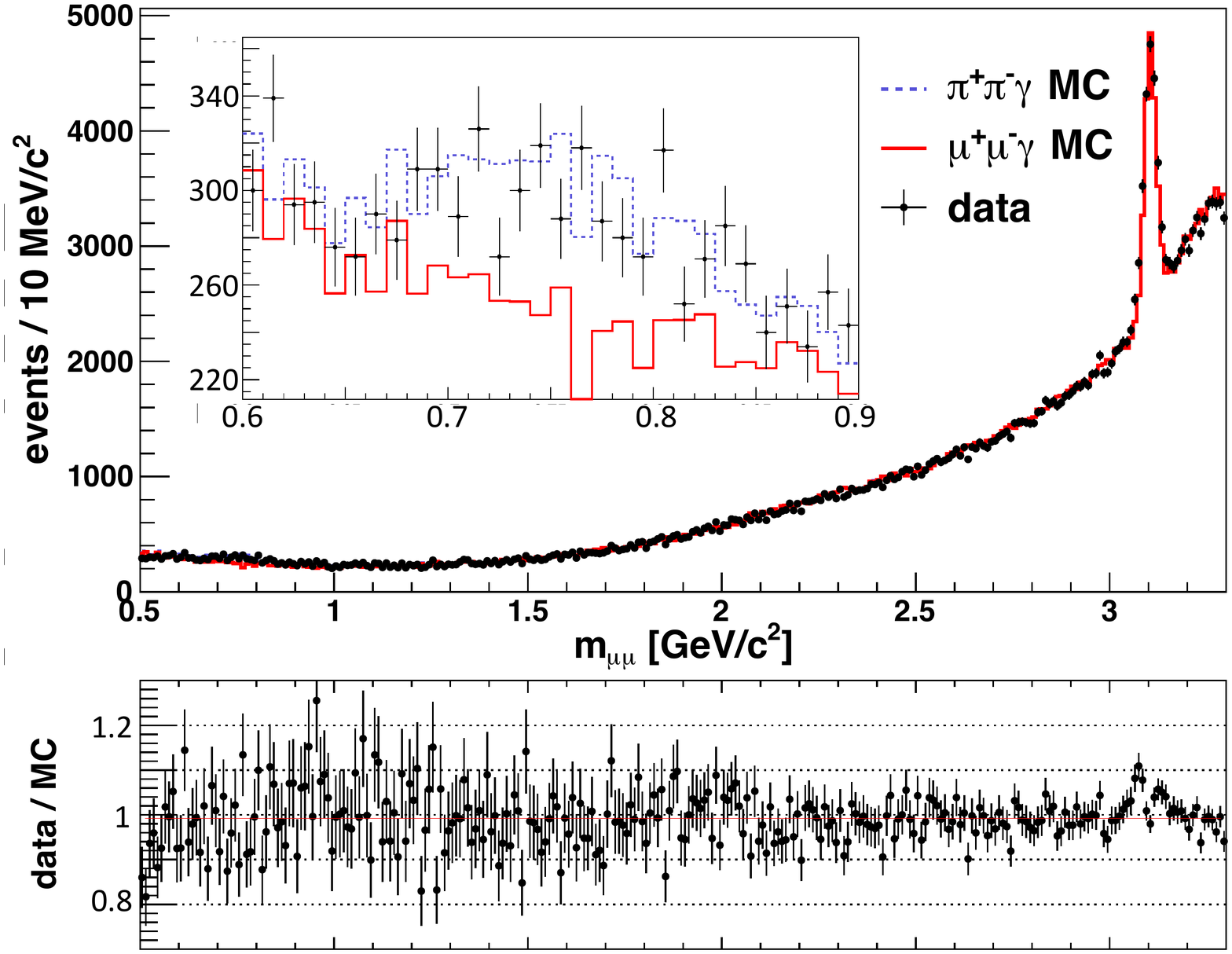}
\caption{Invariant $\mu^+\mu^-$ mass spectrum of data and $\mu^+\mu^-\gamma$ MC after using the ANN as muon selector and applying the efficiency corrections. The upper panel presents the event yield found in data and MC. The inlay shows the zoom for invariant masses between 0.6 and 0.9 GeV/$c^2$. The MC sample is scaled to the luminosity of the data set. The lower panel shows the ratio of these two histograms. A linear fit is performed to quantify the data-MC difference.}
\label{QED_test}
\end{figure}

\noindent We perform a cross check by selecting the well-known $e^+e^-\rightarrow \mu^+\mu^-\gamma$ QED process. This is done using the ANN to select muons instead of pions. Comparing the data to the precise QED prediction of the {\sc phokhara} generator and applying the mentioned efficiency corrections on the MC sample, we find an excellent agreement. Figure \ref{QED_test} shows the corresponding event yield in data and MC. The lower panel shows their relative discrepancy. A linear fit is performed to quantify their difference leading to a discrepancy of ($1.0\pm 0.3\pm0.9$)\% and $\chi^2/{\rm ndf} = 134/139$ over the full mass range.

\section{Extraction of the Cross Section}

\begin{figure}[htb]
\centering
\includegraphics[width = 12cm]{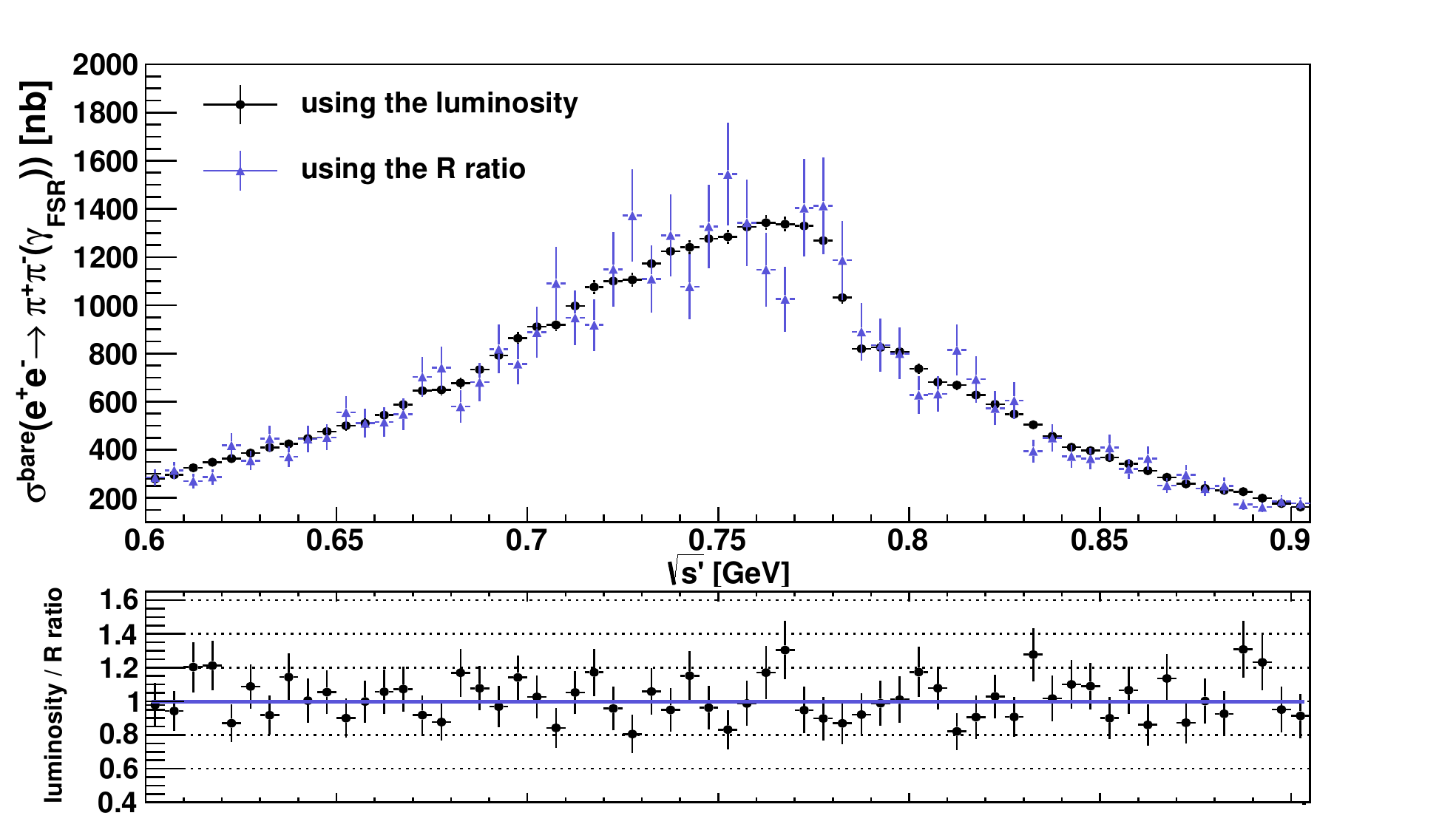}
\caption{Comparison between the methods to extract $\sigma_{\pi\pi}$ as described in the text --- using the luminosity and radiator function (black) and normalizing to the number of ${\mu^+\mu^-\gamma}$ events (blue). The lower panel shows the ratio of these results together with a linear fit (blue line) to quantify their difference, which is found to be (0.85 $\pm$ 1.68)\% and $\chi^2/{\rm ndf} = 50/60$, where the error is statistical.}
\label{comparison}
\end{figure}

Two independent normalization schemes can be applied to extract $\sigma_{\pi\pi}$. On the one hand, one can normalize the $\pi^+\pi^-\gamma$ event yield $N_{\pi\pi\gamma}$ to the luminosity of the used data set, the selection efficiency determined with signal MC samples from the {\sc phokhara} generator, and the theoretical radiator function \cite{radiator}, as it has been done in \cite{2pi_KLOE08,2pi_KLOE10}. On the other hand, one can calculate the R ratio, or in other words, one normalizes $N_{\pi\pi\gamma}$ to the number of $\mu^+\mu^-\gamma$ events, as it has been done in \cite{2pi_KLOE12,2pi_BaBar_PRL,2pi_BaBar}. Both methods have been applied in the analysis, and agree within the errors, as presented in Fig. \ref{comparison}. The first method described above is used to obtain the final result of this analysis, since the second one is limited by the $\mu^+\mu^-\gamma$ statistics.\\

\noindent As input for the calculation of $a_\mu^{\pi\pi}$ the bare cross section is needed, which is in addition corrected for final state radiation (FSR), $\sigma^{\rm bare}(e^+e^-\rightarrow\pi^+\pi^-(\gamma_{\rm FSR}))$. It can be obtained by dividing the cross section $\sigma_{\pi\pi}$ by  the vacuum polarization correction obtained with the {\sc phokhara} generator. As pointed out in Ref.~\cite{Jegerlehner}, in order to consider radiative effects in the dispersion integral for $a_\mu$, an FSR correction has to be performed. The correction factor is determined with the {\sc phokhara} generator in next-to-leading order and a theoretical correction factor taken from \cite{FSR_Schwinger}.\\

\noindent More detailed information about the extraction methods and applied corrections can be found in section 6 of Ref. \cite{2pi_BES}.

\section{Results}

\begin{figure}[htb]
\centering
\includegraphics[width = 7.5cm]{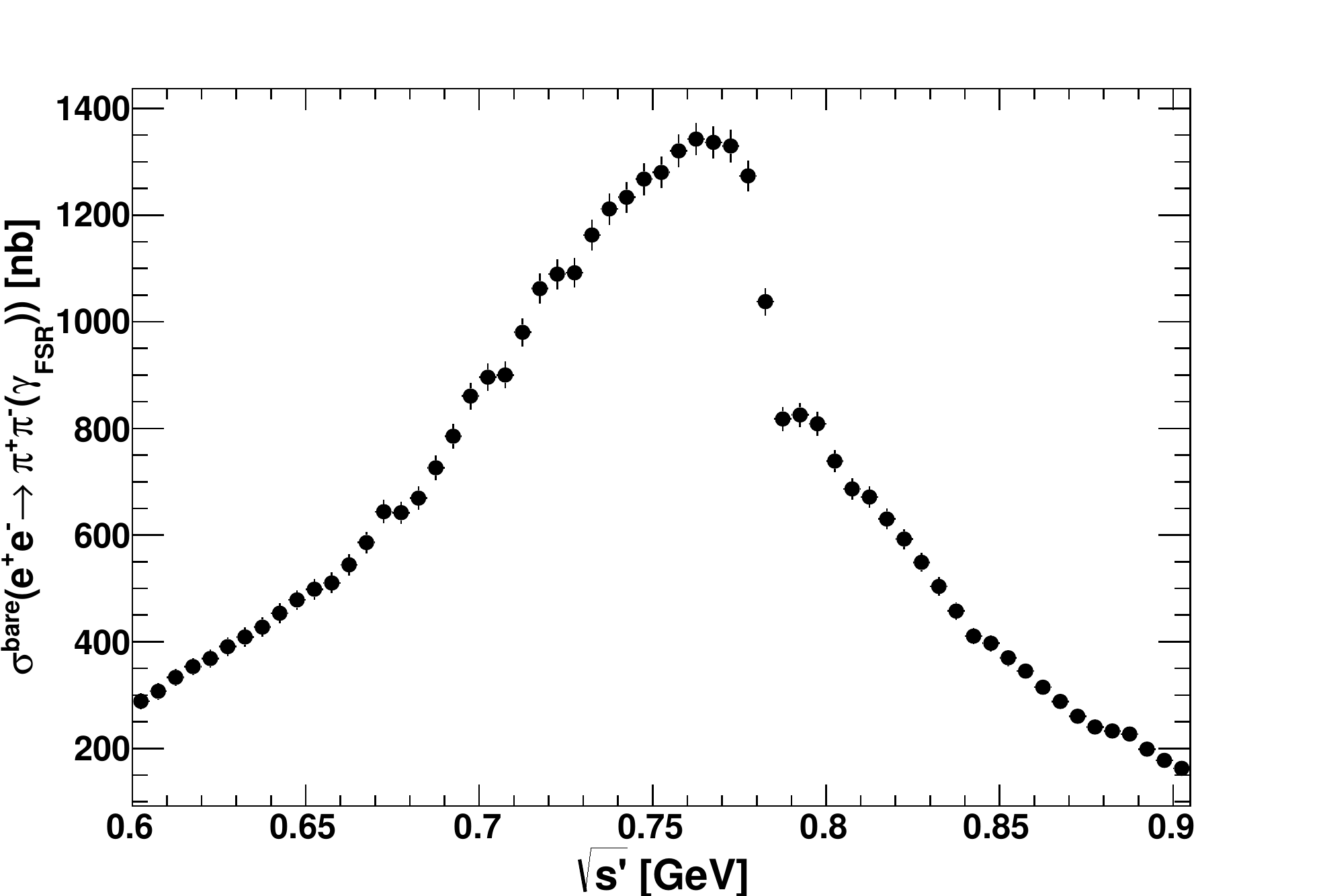}
\includegraphics[width = 7.5cm]{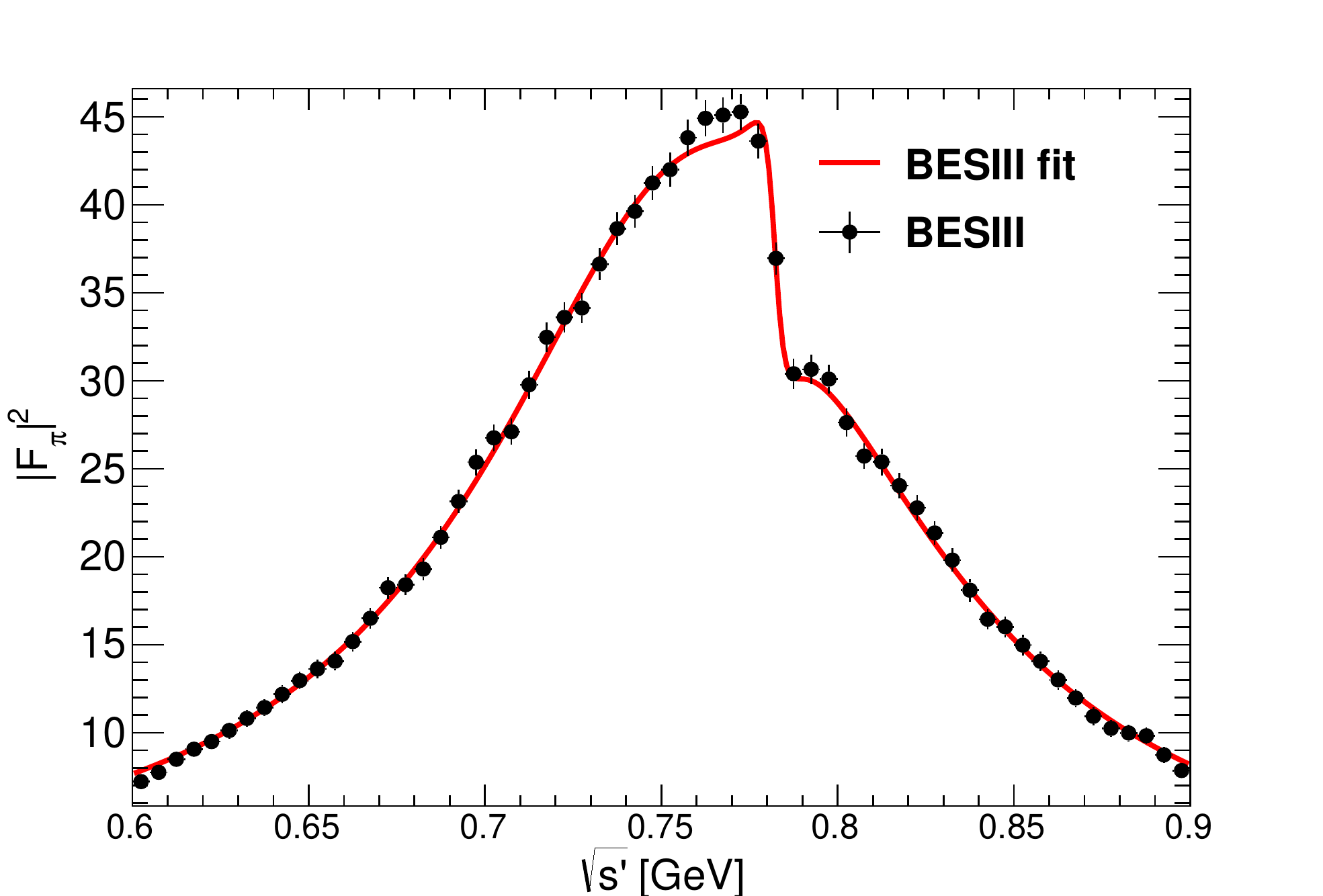}
\caption{Left: The measured bare $e^+e^-\rightarrow\pi^+\pi^-(\gamma_{\rm FSR})$ cross section, showing statistical errors only.
Right: The extracted pion form factor, which is fitted with the Gounaris-Sakurai parametrization as described in the text (solid line).}
\label{result_crossSection}
\end{figure}

The result for $\sigma^{\rm bare}(e^+e^-\rightarrow\pi^+\pi^-(\gamma_{\rm FSR}))$ as a function of $\sqrt{s}=m_{\pi\pi}$ is shown in Fig.~\ref{result_crossSection}. The cross section is corrected for vacuum polarization effects, which are taken from the {\sc phokhara} event generator \cite{Phokhara7}, and includes final state radiation. An unfolding procedure using the Singular Value Decomposition method \cite{unfolding} is applied. The well-known structure of the $\rho$-$\omega$ interference can be observed in the $\rho(770)$ peak region. \\

\noindent The result for the pion form factor $|F_\pi|^2$ is also shown in Fig.~\ref{result_crossSection}. Vacuum polarization corrections are applied, but, differently from the cross section shown in Fig.~\ref{result_crossSection}, final state radiation effects are excluded here. The numerical values of the cross section and the form factor can be found in Ref.  \cite{2pi_BES}. The solid line in Fig.~\ref{result_crossSection} represents a fit to data using the Gounaris and Sakurai \cite{GS} parametrization. It is in excellent agreement with the BESIII data in the full mass range from 600 to 900 MeV/$c^2$, resulting in $\chi^2/{\rm ndf} = 49.1/56$. 
The mass and width of the $\rho$ meson, the mass of the $\omega$ meson, and the phase of the Breit-Wigner function are free parameters. The width of the $\omega$ meson is fixed to the PDG value \cite{PDG2014}. 
Corresponding amplitudes for the higher $\rho$ states, $\rho'$, $\rho''$, and $\rho'''$, as well as the masses and widths of those states were taken from Ref.~\cite{2pi_BaBar}.\\

\begin{figure}[htb]
\centering
\includegraphics[width = 7.5cm]{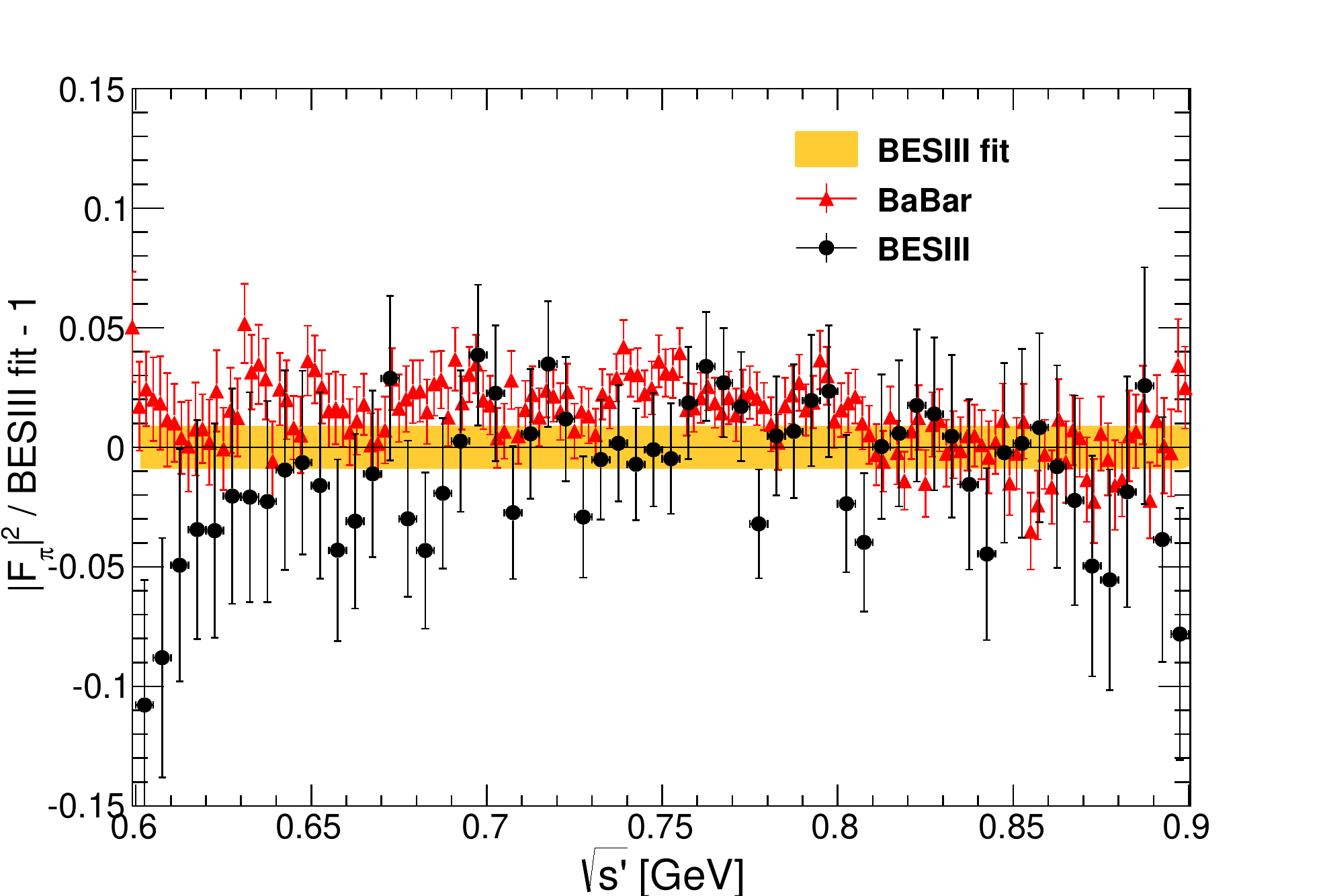}
\includegraphics[width = 7.5cm]{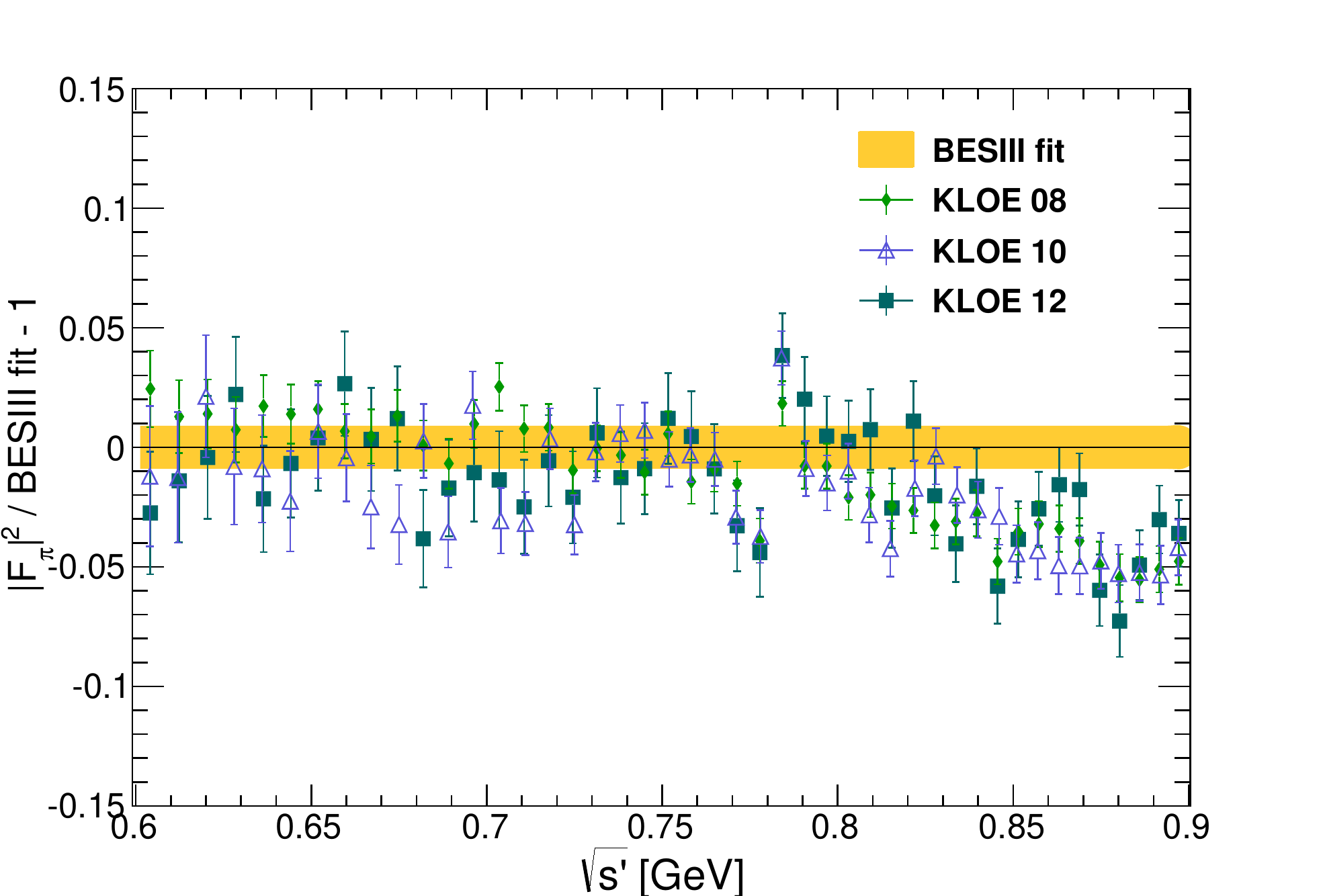}
\caption{Relative difference of the form factor squared from BaBar \cite{2pi_BaBar} (left), KLOE \cite{2pi_KLOE08,2pi_KLOE10,2pi_KLOE12} (right), and the BESIII fit (shaded band). Statistical and systematic uncertainties are included in the data points. The width of the BESIII band shows the systematic uncertainty only.}
\label{result_FF}
\end{figure}

\noindent Figure \ref{result_FF} shows a normalization of the BESIII, BaBar \cite{2pi_BaBar}, and KLOE  \cite{2pi_KLOE08,2pi_KLOE10,2pi_KLOE12} pion form factor data to our BESIII fit. We observe a good agreement with the KLOE 08 and KLOE 12 data sets up to the mass range of the $\rho$-$\omega$ interference. In the same mass range the BaBar and KLOE 10 data sets show a systematic shift, the deviation is, however, not exceeding 1 to 2 standard deviations.
At higher masses the agreement with BaBar is very good, while all three KLOE data sets show a discrepancy with BESIII, which is increasing with mass and which is reaching approximately 5\% at 900 MeV/c$^2$.. A comparison with CMD2~\cite{2pi_CMD2_2004,2pi_CMD2_2006}, and SND \cite{2pi_SND} is less conclusive in the $\rho$ tail regions, since the error bars are large. The spectra from SND and from the 2006 publication of CMD-2 are found to be in very good agreement with BESIII in the $\rho$ peak region, while the 2004 result of CMD-2 shows a systematic deviation of a few percent.\\

\begin{figure}[h]
\centering
\includegraphics[width = 14cm]{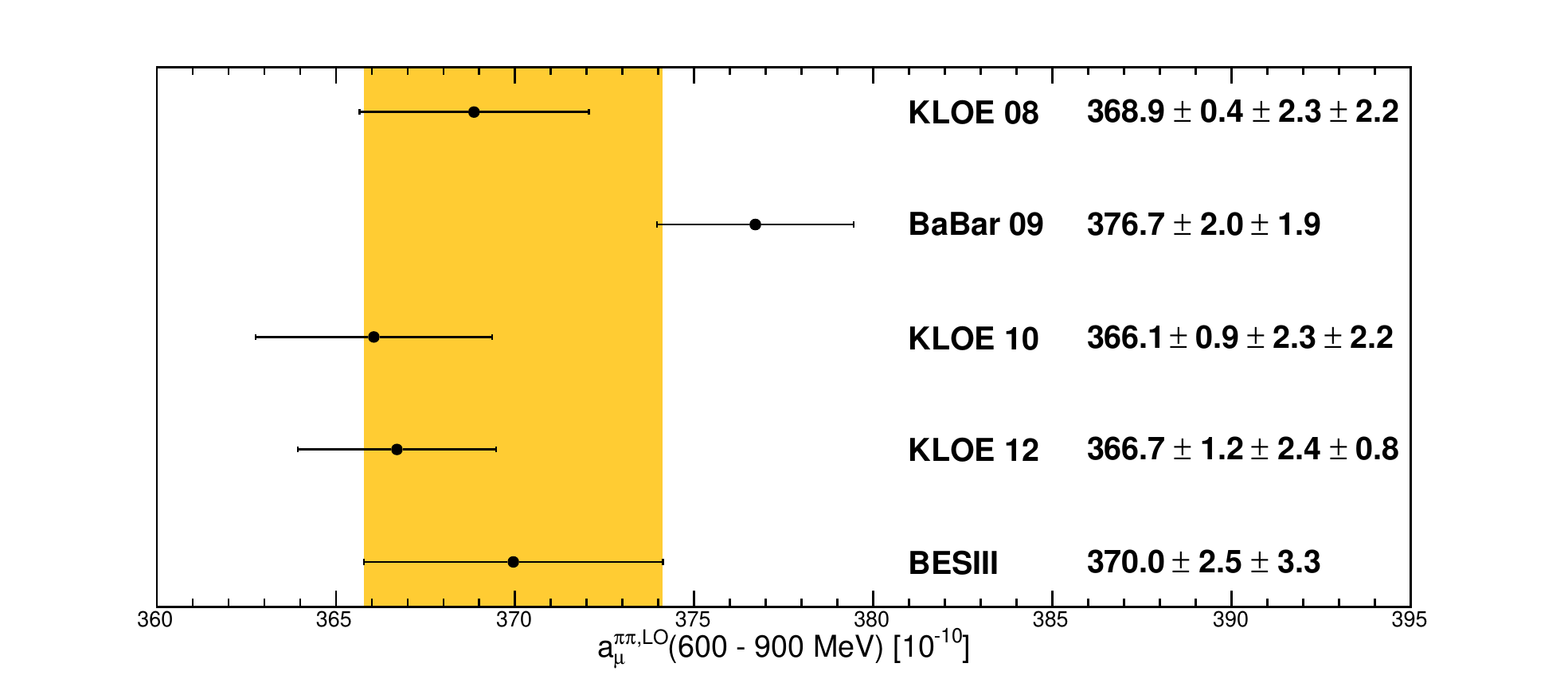}
\caption{Our calculation of the leading-order (LO) hadronic vacuum polarization 2$\pi$ contributions to $(g-2)_\mu$ in the energy range 600 - 900 MeV from BESIII, using the vacuum polarization correction implemented in the {\sc phokhara} generator, and
based on the data sets from KLOE 08 \cite{2pi_KLOE08}, 10 \cite{2pi_KLOE10}, 12 \cite{2pi_KLOE12}, and BaBar \cite{2pi_BaBar}, with the statistical and systematic errors. The statistical and systematic errors are added quadratically. The shaded area shows the 1$\sigma$ range of the BESIII result.}
\label{a_mu}
\end{figure}

\noindent We also calculate the contribution of our cross section measurement to the hadronic contribution of $(g-2)_\mu$ in the energy range between 0.6 and 0.9 GeV,
\small
\begin{equation}
	a_\mu^{\pi\pi,\rm LO}(0.6 - 0.9\, \rm{GeV}) = \frac{1}{4\pi^3}\int\limits_{(0.6\rm{GeV})^2}^{(0.9\rm{GeV})^2} ds' K(s') \sigma^{\rm bare}(e^+e^-\rightarrow\pi^+\pi^-(\gamma_{\rm FSR})) \; ,
\end{equation}
\normalsize where $K(s')$ is the kernel function \cite[Eq. (5)]{Jegerlehner}.
As summarized in Fig.~\ref{a_mu}, the BESIII result, $a_\mu^{\pi\pi,\rm LO}(600-900\;\rm MeV) = (370.0 \pm 2.5_{\rm stat} \pm 3.3_{\rm sys})\cdot 10^{-10}$, using the vacuum polarization correction implemented in the {\sc phokhara} generator, is found to be in between the corresponding values of KLOE and BaBar, calculated based on their published results in the same energy range.

\section{Summary}
We performed a new measurement of the $\sigma^{\rm bare}(e^+e^-\rightarrow{\pi^+\pi^-(\gamma_{\rm FSR})})$ cross section and the pion form factor with an accuracy of 0.9\% in the dominant $\rho(770)$ mass region between 600 and 900 MeV/$c^{2}$ at BESIII. Therefore, we exploited the ISR method. We computed the two-pion contribution to the hadronic vacuum polarization contribution to $(g-2)_\mu$ from the BESIII data to be $a_\mu^{\pi\pi,\rm LO}(600-900\;\rm MeV) = (370.0 \pm 2.5_{\rm stat} \pm 3.3_{\rm sys})\cdot 10^{-10}$, using the vacuum polarization correction implemented in the {\sc phokhara} generator.  It is found to be in the middle between the corresponding values using KLOE or BaBar data in the same energy range.

\end{document}